# Set-membership improved normalized subband adaptive filter algorithms for acoustic echo cancellation

Yi Yu[1], Haiquan Zhao[1*], Badong Chen[2]

**Abstract.** In order to improve the performances of recently-presented improved normalized subband adaptive filter (INSAF) and proportionate INSAF algorithms for highly noisy system, this paper proposes their set-membership versions by exploiting the theory of set-membership filtering. Apart from obtaining smaller steady-state error, the proposed algorithms significantly reduce the overall computational complexity. In addition, to further improve the steady-state performance for the algorithms, their smooth variants are developed by using the smoothed absolute subband output errors to update the step sizes. Simulation results in the context of acoustic echo cancellation have demonstrated the superiority of the proposed algorithms.

## 1. Introduction

In modern communication network, hands-free telephony and teleconference, the acoustic echo is a common problem that must be eliminated to improve the call quality. In recent decades, therefore, acoustic echo cancellation (AEC) based on adaptive filter has obtained significant attention [1]-[3]. Naturally, AEC is also a system identification problem (i.e., identifying the impulse response of the echo path), but it imposes several characteristics on the adaptive filter. Namely, 1) the input is the speech signal which is nonstationary and highly colored, and 2) the impulse response of the echo path is long and sparse. Owing to its simplicity and ease of implementation, the normalized least mean square (NLMS) algorithm is often used in AEC. However, the NLMS algorithm will suffer from slow convergence when the input signal is colored.

To overcome this drawback, the affine projection (AP) family uses $K$ past input signal vectors for updating the filter' coefficients, at each iteration, where $K$ is the projection order [4], [5]. The convergence rate of AP increases as $K$ increases, but meanwhile the steady-state error and computational cost of that become large. To reduce the computational complexity, several AP algorithms with variable projection order [6] as well as fast AP algorithms [7] were opened.

[1] School of Electrical Engineering at Southwest Jiaotong University, Chengdu, 610031, China.

[2] School of Electronic and Information Engineering, Xi'an Jiaotong University, Xi'an, China.

E-mail: yuyi_xyuan@163.com (Y. Yu), hqzhao_swjtu@126.com (H. Zhao), chenbd@mail.xjtu.edu.cn (B. Chen).

* Corresponding author

Alternatively, to deal with the colored input signals, various subband adaptive filters (SAFs) have been proposed [8]-[10]. In all the SAFs, the colored input signal is partitioned into almost mutually exclusive multiple subband signals by the analysis filters; then, the decimated subband signals approximated to white signals are used to update the filter's coefficient vector, thus improving the convergence rate. In [10], the normalized SAF (NSAF) algorithm was developed. Compared with the NLMS, the NSAF provides faster convergence rate for the colored input signals. In addition to this benefit, the NSAF still retains almost the same computational complexity as the NLMS, especially for a long filter. Hence, one of the interesting applications for the NSAF is AEC [11]. Following this algorithm, many works from the following aspects have been reported to obtain its improvement in performance (e.g., the convergence rate, steady-state error and computational complexity) [12]-[16]. One way is by employing the time-varying step size instead of the constant step size in the NSAF [12]-[14], whose aim is to overcome the tradeoff between the convergence rate and steady-state error, at the expense of the computational complexity. On the other hand, to reduce the final estimation error of NSAF when identifying a highly noisy system (also called a low signal-noise-ratio (SNR) system), Choi, *et al.* [15] and Ni [16] proposed almost at the same time an improved NSAF (INSAF) which reusing past weight vectors at each iteration, also with a moderate increase in the computational cost. In addition, in many practical applications, the impulse response to be estimated is usually sparse such as the echo path in AEC [2]. The sparse system has the property that only a fraction of coefficients of the impulse response have large magnitude and the rest coefficients is very small or zero. Aiming to such a system, many proportionate algorithms have been developed [2], [17]-[21]. Specially, to improve the convergence rate of SAF in sparse scenarios, the proportionate-family of NSAF was developed by directly extending the proportionate strategies presented in the NLMS, e.g., the resulting improved proportionate NSAF (IPNSAF) algorithm [19]-[20]. Furthermore, the improved proportionate version of INSAF (IP-INSAF) was proposed for identifying sparse system in highly noisy environments [21].

For adaptive filtering algorithms, it is desired that efforts make sense in pursuing fast convergence rate, low steady-state error and low computational complexity simultaneously. To this end, in [22], Gollamudi *et al.* introduced firstly the set-membership filtering theory into the NLMS, developing the SM-NLMS algorithm. The fundamental idea is that the filter's weight vector is updated only when the magnitude of the output error exceeds a predetermined bound. Benefited from the success of the SM-NLMS, the set-membership filtering has also been applied to other type of adaptive algorithms such as the AP-family algorithm [23]-[26], and frequency-domain NLMS algorithm [27]. Recently, the set-membership strategy has also been investigated in the NSAF and IPNSAF algorithms, respectively, yielding the SM-NSAF [19] and SM-IPNSAF [20] algorithms which work well. As far as we know, however, the behavior of the set-membership method has not been studied in low SNR environments. In this paper, our contributions are as follows:

1) We propose to extend the set-membership filtering to the INSAF and IP-INSAF algorithms, resulting in the SM-INSAF and IP-INSAF algorithms, respectively. The proposed SM-INSAF achieves smaller steady-state error and reduces the overall computational burden as compared to the INSAF. And, the proposed SM-IP-INSAF is more suitable than the proposed SM-INSAF for sparse systems since the former is a proportionate algorithm.

2) Then, a smooth estimation method is proposed for the absolute subband errors used for computing the step sizes in the SM-INSAF and SM-IP-INSAF algorithms, which makes the step sizes decrease smoothly. Hence, the resulting smooth versions have better steady-state performance.

3) What's more, we derive a convergence condition for the existing INSAF and IP-INSAF algorithms. Such a condition also illustrates the stable convergence of the proposed algorithms.

4) Extensive simulations are presented in the context of AEC with a low SNR scenario, which demonstrates the advantages of our proposed algorithms.

The organization of this paper is summarized. Section II describes the original INSAF and IP-INSAF algorithms. In Section III, we develop the set-membership versions of these two algorithms. In Section IV, simulation results are presented to demonstrate the proposed algorithms. Finally, Section V draws conclusions.

## 2. Review of the INSAF and IP-INSAF algorithms

In adaptive AEC, the signal $d(n)$ picked up by the microphone is given by

$$d(n) = \mathbf{u}^T(n)\mathbf{w}_o + v(n), \qquad (1)$$

where $(\cdot)^T$ indicates the transpose of a vector or matrix, $\mathbf{w}_o$ is the impulse response of the acoustic echo path, the echo is generated when the far-end signal $u(n)$ broadcasted by the loudspeaker activates it, $\mathbf{u}(n) = [u(n), u(n-1), ..., u(n-M+1)]^T$ is the input vector, and $v(n)$ is the near-end signal. The near-end signal $v(n)$ may contains the background noise $\eta(n)$ and the near-end speech signal $\vartheta(n)$. The fundamental goal of adaptive echo cancellation is to estimate the echo path $\mathbf{w}_o$ with an adaptive filter $\mathbf{w}(n)$ so that the far-end signal $u(n)$ is filtered by an adaptive filter to produce a replica of the echo $\hat{y}(n)$, $\hat{y}(n) = \mathbf{u}^T(n)\mathbf{w}(n)$. Then, subtracting $\hat{y}(n)$ from $d(n)$, the resulting residual signal, $e(n) = d(n) - \hat{y}(n)$, will be free of echo and will contain only the signal of interest. It is worthy to note that here the near-end speech signal is assumed to be absent, i.e., $\vartheta(n) = 0$, because the adaptation of the echo canceller is usually halted when double-talk situations are detected [17], [18], [28].

For applying SAF to AEC scenario, we prefer to employ the delayless open-loop structure [11] depicted in Fig. 1, since it has no any delay for computing the residual signal $e(n)$ while the original structure in [10] has the signal path delay for reconstructing $e(n)$ by the synthesis filter bank. As shown in Fig. 1, the signals $d(n)$ and $u(n)$ are divided into multiple subband signals $d_i(n)$ and $u_i(n)$, respectively, through the analysis filter bank $H_i(z)$, $i = 0,1,...,N-1$. The subband signals $y_i(n)$ and $d_i(n)$ are separately N-fold decimated to yield $y_{i,D}(k)$ and $d_{i,D}(k)$ with a lower sample rate, i.e., $d_{i,D}(k) = d_i(kN)$ and $y_{i,D}(k) = y_i(kN)$. Note that, $n$ indicates here the index in the original sequence, and $k$ indicates the index in the decimated sequence. The decimated output signal of the $i^{th}$ subband can be formulated as $y_{i,D}(k) = \mathbf{u}_i^T(k)\mathbf{w}(k)$, where $\mathbf{w}(k) = [w_1(k), w_2(k),...,w_M(k)]^T$ denotes the weight vector of adaptive filter at iteration $k$ in the decimated sequence, and $\mathbf{u}_i(k) = [u_i(kN), u_i(kN-1),...,u_i(kN-M+1)]^T$. The $i^{th}$ subband error signal is computed as

$$e_{i,D}(k) = d_{i,D}(k) - y_{i,D}(k). \tag{2}$$

In such a structure, once $\mathbf{w}(k)$ is updated, it will be copied to $\mathbf{w}(n)$ every $N$ input samples.

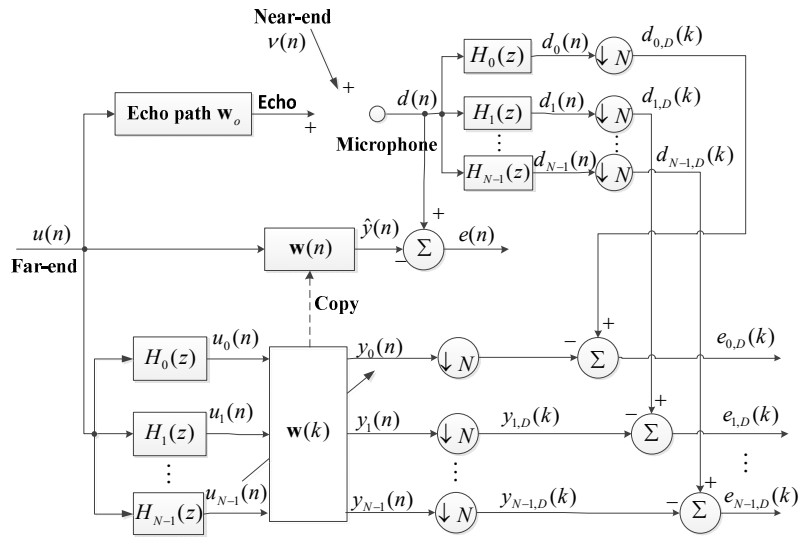

**Fig. 1** Delayless open-loop structure of multiband SAF for AEC.

As in [15], [16], the INSAF algorithm is expressed as

$$\mathbf{w}(k+1) = \frac{1}{P}\sum_{p=0}^{P-1}\mathbf{w}(k-p) + \mu\sum_{i=0}^{N-1}\frac{\xi_{i,D}(k)\mathbf{u}_i(k)}{\|\mathbf{u}_i(k)\|_2^2} \tag{3}$$

where $\mu$ is the step-size, $P$ denotes the number of using past weight vectors, $\|\cdot\|_2$ denotes the $l_2$-norm of a vector, and

$$\xi_{i,D}(k) = d_{i,D}(k) - \mathbf{u}_i^T(k)\frac{1}{P}\sum_{p=0}^{P-1}\mathbf{w}(k-p). \tag{4}$$

For the proportionate INSAF algorithms, the update equation of the weight vector is given by [21]

$$\mathbf{w}(k+1) = \frac{1}{P}\sum_{p=0}^{P-1}\mathbf{w}(k-p) + \mu\sum_{i=0}^{N-1}\frac{\mathbf{G}(k)\mathbf{u}_i(k)\xi_{i,D}(k)}{\mathbf{u}_i^T(k)\mathbf{G}(k)\mathbf{u}_i(k)} \tag{5}$$

where $\mathbf{G}(k) = \text{diag}\{g_1(k), g_2(k),...,g_M(k)\}$ is an $M \times M$ diagonal matrix whose role is to assign an individual gain $g_m(k)$ to every filter weight $w_m(k)$, thereby improving the convergence rate when identifying sparse system. In the IP-INSAF algorithm [20], $g_m(k)$ for $m = 1, 2,..., M$ are calculated as

$$g_m(k) = \frac{1-\lambda}{2M} + (1+\lambda)\frac{|w_m(k)|}{2\|\mathbf{w}(k)\|_1 + \zeta}, \qquad (6)$$

where $\|\cdot\|_1$ indicates the $l_1$-norm of a vector, $\zeta$ is a small positive constant to avoid the division by zero, and $\lambda$ locates in the interval $[-1, 1]$. In practice, good choices of $\lambda$ are $0$ or $-0.5$. It is noteworthy that the formula (6) is firstly designed in the improved PNLMS algorithm [18] which is robust regardless of sparse or dispersive impulse response.

## 3. Proposed set-membership algorithms

### 3.1. Derivation of SM-INSAF

Define the subband input matrix and subband desired signal vector as follows:

$$\mathbf{U}(k) = [\mathbf{u}_0(k), \mathbf{u}_1(k), ..., \mathbf{u}_{N-1}(k)], \qquad (7)$$

$$\mathbf{d}_D(k) = [d_{0,D}(k), d_{1,D}(k), ..., d_{N-1,D}(k)]. \qquad (8)$$

To utilize the advantage of the set-membership concept to improve the performance of the INSAF, a new constrained optimization problem is formulated as

$$\min_{\mathbf{w}(k+1)} \left\{ \sum_{p=0}^{P-1} \rho^p \|\mathbf{w}(k+1) - \mathbf{w}(k-p)\|_2^2 \right\} \qquad (9)$$

subject to

$$\mathbf{d}_D(k) - \mathbf{U}^T(k)\mathbf{w}(k+1) = \mathbf{b}(k) \qquad (10)$$

where $0 < \rho \leq 1$ is the forgetting factor, and $\mathbf{b}(k) = [b_0(k), b_1(k), ..., b_{N-1}(k)]^T$ denotes the subband error-bound vector. Note that when $\mathbf{b}(k) = \mathbf{0}$, (9) and (10) have been used to derive the standard INSAF algorithm [15].

To solve this optimization, we resort to the method of Lagrange multipliers to yield:

$$f(k) = \sum_{p=0}^{P-1} \rho^p \|\mathbf{w}(k+1) - \mathbf{w}(k-p)\|^2 + \left[\mathbf{d}_D(k) - \mathbf{U}^T(k)\mathbf{w}(k+1) - \mathbf{b}(k)\right]^T \boldsymbol{\theta} \qquad (11)$$

where $\boldsymbol{\theta} = [\theta_1, \theta_2, ..., \theta_N]^T$ is the Lagrange multiplier vector. Let the derivative of (11) with respect to $\mathbf{w}(k+1)$ equal to zero, we have

$$\mathbf{w}(k+1) = \alpha \sum_{p=0}^{P-1} \rho^p \mathbf{w}(k-p) + \frac{1}{2}\alpha \mathbf{U}(k)\boldsymbol{\theta} \qquad (12)$$

where $\alpha = \left(\sum_{p=0}^{P-1} \rho^p\right)^{-1}$. Substituting (12) into (10), we get:

$$\boldsymbol{\theta} = 2\alpha^{-1}\left(\mathbf{U}^T(k)\mathbf{U}(k)\right)^{-1}\left(\boldsymbol{\varepsilon}_D(k) - \mathbf{b}(k)\right) \qquad (13)$$

where

$$\boldsymbol{\varepsilon}_D(k) = \mathbf{d}_D(k) - \alpha \mathbf{U}^T(k) \sum_{p=0}^{P-1} \rho^p \mathbf{w}(k-p). \tag{14}$$

Combining (12) and (13), we obtain

$$\mathbf{w}(k+1) = \alpha \sum_{p=0}^{P-1} \rho^p \mathbf{w}(k-p) + \mathbf{U}(k)\left(\mathbf{U}^T(k)\mathbf{U}(k)\right)^{-1}\left(\boldsymbol{\varepsilon}_D(k) - \mathbf{b}(k)\right). \tag{15}$$

Applying the *diagonal assumption* that the off-diagonal elements of the matrix $\mathbf{U}^T(k)\mathbf{U}(k)$ are negligible, which has been frequently used to design subband algorithms [10], [12]-[16], (15) is simplified as

$$\mathbf{w}(k+1) = \overline{\mathbf{w}}(k) + \mu \sum_{i=0}^{N-1} \frac{\left(\varepsilon_{i,D}(k) - b_i(k)\right)\mathbf{u}_i(k)}{\|\mathbf{u}_i(k)\|_2^2} \tag{16}$$

where $\overline{\mathbf{w}}(k) = \alpha \sum_{p=0}^{P-1} \rho^p \mathbf{w}(k-p)$ denotes the weighted-average of $P$ recent weight vectors, and $\varepsilon_{i,D}(k)$ is the $i^{th}$ element of $\boldsymbol{\varepsilon}_D(k)$ expressed as:

$$\varepsilon_{i,D}(k) = d_{i,D}(k) - \mathbf{u}_i^T(k)\overline{\mathbf{w}}(k). \tag{17}$$

According to the set-membership filtering theory [22]-[26], [29], one know that if the average weight vector $\overline{\mathbf{w}}(k)$ lies in the constraint set $\mathrm{H}_k$ at $k^{th}$ iteration defined by

$$\mathrm{H}_k = \left\{\overline{\mathbf{w}} : \left|d_{i,D}(k) - \mathbf{u}_i^T(k)\overline{\mathbf{w}}\right| \leq \gamma_i, \; i=0,1,...,N-1\right\}, \tag{18}$$

then the adaptive filter will stop updating and the adaptive solution is obtained, where $\gamma_i$ is a prespecified boundary related to the $i^{th}$ subband error to determine the constraint set $\mathrm{H}_k$. In other words, only when the condition $|\varepsilon_{i,D}(k)| > \gamma_i$ for the $i^{th}$ subband is satisfied, the corresponding subband signals $\mathbf{u}_i(k)$ and $\varepsilon_{i,D}(k)$ take part in the update of the weight vector. Hence, (16) can be changed as

$$\mathbf{w}(k+1) = \overline{\mathbf{w}}(k) + \mu \sum_{i \in i_s} \frac{\left(\varepsilon_{i,D}(k) - b_i(k)\right)\mathbf{u}_i(k)}{\|\mathbf{u}_i(k)\|_2^2} \tag{19}$$

where $i_s = \left\{i=0,1,...,N-1 \big| |\varepsilon_{i,D}(k)| > \gamma_i\right\}$ is a set of subband index corresponding to the condition $|\varepsilon_{i,D}(k)| > \gamma_i$. Apparently, the cardinality of the set $i_s$ is less than or equal to $N$.

Next, an important problem is how to choose the subband error-bounds $b_i(k), i=0,1,...,N-1$. It has been reported in [22] that the only condition of choosing $b_i(k)$ is $|b_i(k)| \leq \gamma_i$. Evidently, for such a condition, there are an infinite number of ways for choosing $b_i(k)$, and each one can lead to a different algorithm. However, a widely used way is to choose $b_i(k) = \gamma_i \text{sgn}(\varepsilon_{i,\text{D}}(k))$ [19]-[20], [22] so that the adaptive solution locates at the nearest boundary of $H_k$, where $\text{sgn}(\cdot)$ indicates the sign function. Applying such a choice into (19), the SM-INSAF algorithm for updating the weight vector is expressed as

$$\mathbf{w}(k+1) = \bar{\mathbf{w}}(k) + \sum_{i=0}^{N-1} \mu_i(k) \frac{\varepsilon_{i,\text{D}}(k)\mathbf{u}_i(k)}{\|\mathbf{u}_i(k)\|_2^2} \qquad (20)$$

where the step size $\mu_i(k)$:

$$\mu_i(k) = \begin{cases} 1 - \dfrac{\gamma_i}{|\varepsilon_{i,\text{D}}(k)|}, & \text{if } |\varepsilon_{i,\text{D}}(k)| > \gamma_i \\ 0, & \text{otherwise} \end{cases}. \qquad (21)$$

Note that, in the case of $\rho = 1$, the SM-INSAF algorithm reduces to

$$\mathbf{w}(k+1) = \frac{1}{P}\sum_{p=0}^{P-1}\mathbf{w}(k-p) + \sum_{i=0}^{N-1} \mu_i(k) \frac{\varepsilon_{i,\text{D}}(k)\mathbf{u}_i(k)}{\|\mathbf{u}_i(k)\|_2^2} \qquad (22)$$

where

$$\varepsilon_{i,\text{D}}(k) = \xi_{i,\text{D}}(k) = d_{i,\text{D}}(k) - \mathbf{u}_i^T(k)\frac{1}{P}\sum_{p=0}^{P-1}\mathbf{w}(k-p). \qquad (23)$$

### 3.2. Derivation of SSM-INSAF

To further improve the steady-state performance of the SM-INSAF, our motivation is to replace $|\varepsilon_{i,\text{D}}(k)|$, $i = 0,1,...,N-1$, in (21) with their own smooth estimates $\sigma_{\varepsilon_{i,\text{D}}}(k)$, i.e.,

$$\sigma_{\varepsilon_{i,\text{D}}}(k) = \beta \sigma_{\varepsilon_{i,\text{D}}}(k-1) + (1-\beta)|\varepsilon_{i,\text{D}}(k)| \qquad (24)$$

where $\beta$ is the smooth factor whose value is smaller than but close to 1, generally, it can be chosen by $\beta = 1 - N/\kappa M$, where $\kappa \geq 1$ [14], [31], and the initial value of $\sigma_{\varepsilon_{i,\text{D}}}(k)$ is zero. Accordingly, (21) is changed as

$$\mu_i(k) = \begin{cases} 1 - \dfrac{\gamma_i}{\sigma_{\varepsilon_{i,\text{D}}}(k)}, & \text{if } |\varepsilon_{i,\text{D}}(k)| > \gamma_i \\ 0, & \text{otherwise} \end{cases}. \qquad (25)$$

It is remarkable that $\sigma_{\varepsilon_{i,\text{D}}}(k) < \gamma_i$ is possible according to (24) so that the value of $\mu_i(k)$ from (25) could be negative. Thus, to avoid this problem, a practical consideration is to use:

$$\mu_i(k) = \begin{cases} 1 - \dfrac{\gamma_i}{\sigma_{\varepsilon_{i,\text{D}}}(k)}, & \text{if } \min(|\varepsilon_{i,\text{D}}(k)|, \sigma_{\varepsilon_{i,\text{D}}}(k)) > \gamma_i \\ 0, & \text{otherwise} \end{cases}. \qquad (26)$$

Iterating repeatedly (24), we have

$$\sigma_{\varepsilon_{i,\text{D}}}(k) = \beta^k \sigma_{\varepsilon_{i,\text{D}}}(0) + (1-\beta)\sum_{l=1}^{k} \beta^{k-l} \left|\varepsilon_{i,\text{D}}(l)\right| \qquad (27)$$
$$= (1-\beta)\sum_{l=1}^{k} \beta^{k-l} \left|\varepsilon_{i,\text{D}}(l)\right| \approx E\left\{\left|\varepsilon_{i,\text{D}}(k)\right|\right\}$$

where $E\{\cdot\}$ denotes the expectation of a random variable. Note that in adaptive algorithms, the form (24) is a very practical way to estimate the statistical value of the instantaneous random variable [12]-[14], [31]. It follows that using the statistical estimate $\sigma_{\varepsilon_{i,\text{D}}}(k)$ instead of its instantaneous value $\left|\varepsilon_{i,\text{D}}(k)\right|$ for computing $\mu_i(k)$, mitigates the adverse effect of the fluctuation of $\left|\varepsilon_{i,\text{D}}(k)\right|$. This phenomenon can be observed from Fig. 5 in Section 4.2. By doing so, the resulting SSM-INSAF can reach to a smaller steady-state error in relation to the SM-INSAF. This modification may make that the SSM-INSAF is not in line with the set-membership filtering concept. However, simulations show that the SSM-INSAF still retains the main property of the set-membership algorithm, i.e., reducing the overall computational complexity, see Table 3.

**3.3. Set-membership versions of the IP-INSAF**

In the sequel, the set-membership versions of the IP-INSAF will be obtained by solving the following optimization problem:

$$\min_{\mathbf{w}(k+1)} \left\{ \sum_{p=0}^{P-1} \rho^p \left[\mathbf{w}(k+1) - \mathbf{w}(k-p)\right]^T \mathbf{G}^{-1}(k) \left[\mathbf{w}(k+1) - \mathbf{w}(k-p)\right] \right\} \qquad (28)$$

with the constraint (10), where $\mathbf{G}^{-1}(k)$ stands for the inverse matrix of $\mathbf{G}(k)$. Accordingly, the Lagrange function is constructed as

$$f_{\text{prop}}(k) = \sum_{p=0}^{P-1} \rho^p \left[\mathbf{w}(k+1) - \mathbf{w}(k-p)\right]^T \mathbf{G}^{-1}(k) \left[\mathbf{w}(k+1) - \mathbf{w}(k-p)\right] \\ + \left[\mathbf{d}_D(k) - \mathbf{U}^T(k)\mathbf{w}(k+1) - \mathbf{c}(k)\right]^T \boldsymbol{\theta} \qquad (29)$$

Then, using the same procedures as Section 3.1 for dealing with (29), the following update equation is obtained as:

$$\mathbf{w}(k+1) = \overline{\mathbf{w}}(k) + \sum_{i=0}^{N-1} \mu_i(k) \frac{\mathbf{G}(k)\mathbf{u}_i(k)\varepsilon_{i,\text{D}}(k)}{\mathbf{u}_i^T(k)\mathbf{G}(k)\mathbf{u}_i(k)}. \qquad (30)$$

In the derivation process of (30), we also use a reasonable *diagonal assumption* that the off-diagonal elements of the matrix $\mathbf{U}^T(k)\mathbf{G}(k)\mathbf{U}(k)$ can be neglected as compared to its diagonal elements [21]. As a consequence, the SM-IP-INSAF algorithm is obtained if the step sizes $\mu_i(k)$, $i = 0,1,...,N-1$ in (30) are computed by (21); and if using (26) to compute $\mu_i(k)$, the SSM-IP-INSAF algorithm is obtained. Note that, when $\rho = 1$, (30) is simplified as

$$\mathbf{w}(k+1) = \frac{1}{P}\sum_{p=0}^{P-1}\mathbf{w}(k-p) + \sum_{i=0}^{N-1}\mu_i(k)\frac{\mathbf{G}(k)\mathbf{u}_i(k)\xi_{i,\text{D}}(k)}{\mathbf{u}_i^T(k)\mathbf{G}(k)\mathbf{u}_i(k)}. \qquad (31)$$

### 3.4. Discussions

**Remark 1:** We investigate the relations of the proposed algorithms and some existing algorithms, as follows:

1) If $P$ is equal to 1, the SM-INSAF will become the SM-NSAF in [19], and if $\gamma_i = 0$, the SM-INSAF reduces to the INSAF in [16]. Hence, the proposed SM-INSAF is a generalization of both the SM-NSAF and INSAF algorithms.

2) Likewise, the proposed SM-IP-INSAF reduces to the SM-IPNSAF in [20] when $P=1$ and the IP-INSAF in [21] when $\gamma_i = 0$. In other words, the proposed SM-IP-INSAF is also a generalization of both the SM-IPNSAF and IP-INSAF algorithms.

3) From another viewpoint, the SM-IP-INSAF is a proportionate variant of the SM-INSAF. And, when $\mathbf{G}(k)$ is the identity matrix, both the algorithms are equivalent.

4) In addition, the proposed SSM-INSAF and SSM-IP-INSAF algorithms are modified versions of the SM-INSAF and SM-IP-INSAF algorithms, respectively, by using (26) instead of (21).

**Remark 2:** It has been studied in [15], [16] that the INSAF has better steady-state performance than the NSAF in low SNR cases, while maintaining comparable convergence performance. This is because the determination of $\mathbf{w}(k+1)$ in the INSAF relies on the weighted-average of the current weight vector $\mathbf{w}(k)$ and past weight vectors $\mathbf{w}(k-p)$, $p=1,...,P-1$, which makes the INSAF prevent the weight vector from fluctuating around the optimum solution $\mathbf{w}_o$ especially in considerable noisy cases [15]. As statement in Remark 1, the proposed SM-INSAF is an improvement of the SM-NSAF by means of reusing the past weight vectors at each iteration. Therefore, based on the same mechanism, the performance of the SM-INSAF should also outperform that of the SM-NSAF in the steady-state when working in a low SNR scenario. On the other hand, the SM-INSAF can be considered as a variable step size INSAF algorithm whose time-varying step size is given by (21), thus it can overcome the tradeoff problem of the INSAF between fast convergence rate and small steady-state error. As shown in simulations, the SM-INSAF reduces the steady-state error as compared to the INSAF under the same convergence rate. In addition, since $|\varepsilon_{i,\text{D}}(k)|$ is smoothed by (26) for updating the step size $\mu_i(k)$, the resulting SSM-INSAF further reduces the steady-state error of the SM-INSAF. In the same token, when identifying sparse impulse responses in low SNR cases, the proposed SM-IP-INSAF and SSM-IP-INSAF are superior to their counterparts (i.e., the SM-IPNSAF and IP-INSAF algorithms) in the steady-state error. As shown in (21) and (26), the step sizes $\mu_i(k)$ at each iteration are always in the range of $[0,\ 1)$. Based on the convergence condition $0<\mu<2$ in Appendix, therefore, one can conclude that the proposed algorithms are stable convergence.

It is worth noting that, in high SNR environments, the fluctuation of the weight vector caused by the system noise is negligible since the system noise is very small. Hence in that case, although the algorithms based on reusing the past weight vectors (e.g., the INSAF and proposed algorithms) can also reduce the steady-state error by increasing $P$, their convergence rates are slowed.

**Remark 3:** Similar to the set-membership algorithms in [19]-[20], the boundary parameter $\gamma_i$ can be chosen by $\gamma_i = \sqrt{t\sigma_\eta^2/N}$. Although the proposed algorithms can improve the performance as compared to their original versions, the choice of the parameter $t$ for them must consider a tradeoff among the convergence rate, steady-state error and computational burden. Namely, a large $t$ reduces the steady-state error and overall computational complexity of the proposed algorithms while slows the convergence rate; conversely, a small $t$ improve the convergence rate of the proposed algorithms while increases their steady-state error and overall computational complexity. Nevertheless, based on our extensive simulations, it is found that $t \in [1,\ 4]$ for the SM-INSAF and SM-IP-INSAF algorithms, and $t \in [0.7,\ 0.9]$ for the SSM-INSAF and SSM-IP-INSAF algorithms can achieve good tradeoff performance.

**Remark 4:** For a set-membership algorithm, the computational cost consists of two parts: $C_{\text{up}}$ and $C_{\text{nup}}$, where $C_{\text{up}} \gg C_{\text{nup}}$, $C_{\text{up}}$ accounts for the operations when an update is performed, and $C_{\text{nup}}$ accounts for the operations when no update is performed. Then, the average number of operations per iteration is expressed by [32]

$$C_{\text{av}} = F_{\text{up}}C_{\text{up}} + (1-F_{\text{up}})C_{\text{nup}} \qquad (32)$$

where $F_{\text{up}}$ is called the update rate, with $F_{\text{up}} < 1$ being for the set-membership algorithms (e.g., the proposed algorithms) and $F_{\text{up}} = 1$ being for the non-set-membership algorithms (e.g., the INSAF algorithm). Hence, according to (32), compared with the traditional INSAF and IP-INSAF algorithms, the proposed set-membership versions reduce the computational complexity in the overall adaptation process. Tables 1 and 2 summarize the computational complexity of several adaptive algorithms in terms of $C_{\text{up}}$ and $C_{\text{nup}}$ per each input sample, where the set-membership AP (SM-AP) [23] and set-membership PAPA (SM-PAPA) [25] belong to the family of AP, $O(K^3)$ denotes the amount for calculating the inverse of a $K \times K$ matrix, and $L$ denotes the length of the analysis filter. Note that, the computational complexities for all SAF algorithms are computed under the delayless open-loop structure. As shown in Tables 1 and 2, the computational complexity of these two AP algorithms is large when $K > 2$.

**Table 1.** Computational complexity of various non-proportionate algorithms for each input sample,

| Algorithms | Additions | Multiplications | Divisions | Comparisons |
|---|---|---|---|---|

| | | | | | |
|---|---|---|---|---|---|
| Update ($C_{up}$) | SM-AP | $(K^2+K-1)M$ $+O(K^3)$ | $(2K^2+K+1)M+K^2$ $+O(K^3)$ | 0 | 1 |
| | INSAF | $4M+2N(L-1)+$ $M(P-1)/N$ | $4M+M/N+2NL+1$ | 1 | 0 |
| | SM-INSAF | $4M+2N(L-1)+$ $M(P-1)/N+1$ | $4M+M/N+2NL+1$ | 2 | 1 |
| | SSM-INSAF | $4M+2N(L-1)+$ $M(P-1)/N+2$ | $4M+M/N+2NL+3$ | 2 | 2 |
| No update ($C_{nup}$) | INSAF | - | - | - | - |
| | SM-INSAF and SSM-INSAF | $2M+2N(L-1)+$ $M(P-1)/N$ | $2M+M/N+2NL$ | 0 | 0 |
| | SM-AP | $M+1$ | $M$ | 1 | |

**Table 2.** Computational complexity of various proportionate algorithms for each input sample.

| | Algorithms | Additions | Multiplications | Divisions | Comparisons |
|---|---|---|---|---|---|
| Update ($C_{up}$) | SM-PAPA | $(K^2+K+1)M$ $+O(K^3)$ | $(2K^2+2K+2)M$ $+O(K^3)$ | 1 | 1 |
| | IP-INSAF | $6M+2N(L-1)+$ $M(P+1)/N$ | $6M+3M/N+2NL+1$ | $1+M/N$ | 0 |
| | SM-IP-INSAF | $6M+2N(L-1)+$ $M(P+1)/N+1$ | $6M+3M/N+2NL+1$ | $2+M/N$ | 1 |
| | SSM-IP-INSAF | $6M+2N(L-1)+$ $M(P+1)/N+2$ | $6M+3M/N+2NL+3$ | $2+M/N$ | 2 |
| No update ($C_{nup}$) | IP-INSAF | - | - | - | - |
| | SM-IP-INSAF and SSM-IP-INSAF | $2M+2N(L-1)+$ $M(P-1)/N$ | $2M+M/N+2NL$ | 0 | 0 |
| | SM-PAPA | $M+1$ | $M$ | 1 | |

## 4. Simulation results

To evaluate the performance of the proposed algorithms, simulations are performed in the context of AEC. Either a sparse or a dispersive acoustic echo path $\mathbf{w}_o$ with $M$=512 taps, shown in Fig. 2 [3], needs to be estimated. The cosine modulated filter bank is used in all the SAF algorithms with 60 dB stopband attenuation for its prototype filter [10], [30]. The background noise $\eta(n)$ is a white Gaussian process with a low SNR of 10dB. Here, it is assumed that the variance of the background noise, $\sigma_\eta^2$, is known for all the set-membership algorithms, because it can be easily estimated online in practice like in [12], [32], [33]. The colored input signal $u(n)$ is either an autoregression signal or a speech signal, where the autoregression input, AR(1), is generated by filtering a zero-mean white Gaussian signal through a first-order autoregression system with a pole at 0.9. The normalized mean square deviation (NMSD), $10\log_{10}(\|\mathbf{w}_o - \mathbf{w}(k)\|_2^2 / \|\mathbf{w}_o\|_2^2)$, (dB), is defined as performance measure. All the curves are the average of 100 independent trials.

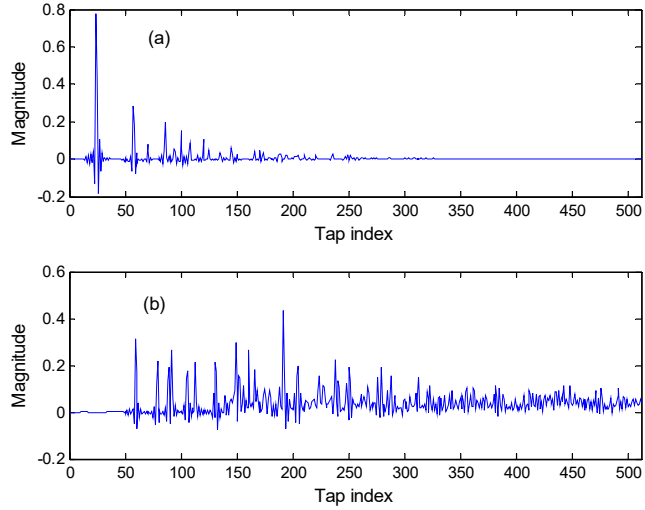

**Fig. 2** Two acoustic echo paths. (a) Sparse case. (b) Dispersive case.

### 4.1. Effect of the parameters *N*, *ρ*, and *P*

In Fig. 3, We investigate the effect of the parameters ($N$, $\rho$, and $P$) one by one on the performance of the SM-INSAF algorithm using AR(1) as the input signal, where the unknown $\mathbf{w}_o$ is depicted in Fig. 2(b). Fig. 3(a) shows the NMSD curves of the algorithm using $N$=2, 4, or 8 subbands. As expected, the algorithm with a large $N$ has faster convergence rate than that with a small $N$. The reason yielding this phenomenon is that each decimated subband input signal is closer to a white signal for a larger $N$. However, for a deterministic colored input signal, when number of subbands is larger than a certain value (e.g., in this case, $N$=8), this phenomenon will not obvious. Fig. 3(b) displays the NMSD curves of the SM-INSAF when $\rho$=0.2, 0.6, or 1. As can be seen from this subfigure, the value of $\rho$ is closer to 1, the steady-state NMSD is smaller while the convergence rate is slightly slow. Fig. 3(c) shows the NMSD curves of the SM-INSAF for $P$=1, 2, or 3. As one can see, a larger $P$ reduces the steady-state NMSD of the SM-INSAF, whilst maintaining comparable convergence performance as the SM-INSAF with $P$=1. Without loss of generality, the results above obtained from Fig. 3 are also reasonable for the proposed SSM-INSAF, SM-IP-INSAF and SSM-IP-INSAF algorithms. In the following examples, therefore, for a fair comparison we set $N$=8 for all SAF algorithms, $P$=2 for the INSAF and IP-INSAF algorithms, and $P$=2, $\rho$=1 for four proposed algorithms.

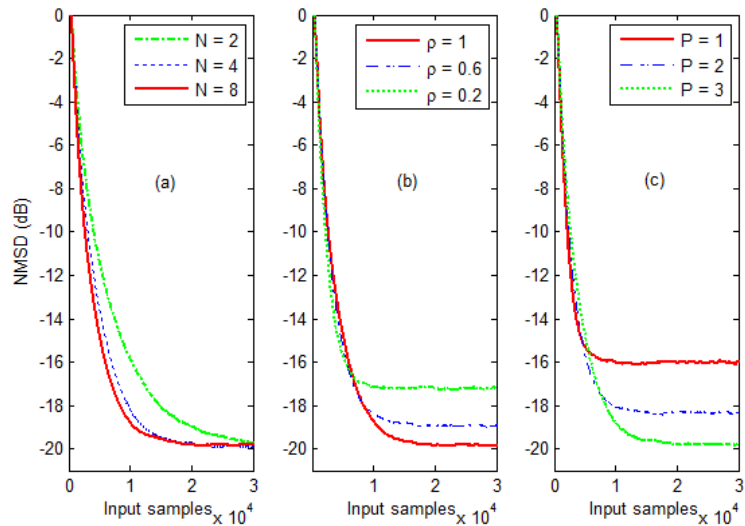

**Fig.** 3. The NMSD curves of the SM-INSAF versus various parameters. (a) different *N* [other parameters: *P*=3, $\rho = 1$, and *t*=2], (b) different *ρ* [other parameters: *P*=3, *N*=8 and *t*=2], (c) different *P* [other parameters: $\rho = 1$, *N*=8 and *t*=2.

### 4.2. Comparisons of subband algorithms

Note that, in order to compare the tracking capability of the algorithms, an abrupt change of the echo path occurs at the middle of input samples, by shifting the impulse response to the right by 12 samples [12].

First, we compare the NMSD performance of the proposed SM-INSAF and SSM-INSAF algorithms with that of the NSAF [10], INSAF with two step sizes ($\mu = 1$ and 0.1) [16] and SM-NSAF [19] algorithms for a dispersive echo path given by Fig. 2(b), in Fig. 4. When using the speech input signal (here, a female speech in English, which is a nonstationary and highly colored signal), to avoid the division by zero in the update formulas of these algorithms, a regularization additive number with the value of $\delta = \sigma_u^2$ is added in their denominators, where $\sigma_u^2$ is the power of the input signal. As can be seen, in comparison with the NSAF, the INSAF works better in the steady-state performance for such a low SNR case since it use past weight vectors in adaptation process. Also, the set-membership algorithms (e.g., the SM-NSAF) achieve a good trade-off between the steady-state NMSD and convergence rate in contrast to their counterparts (e.g., the NSAF). This is due to the fact that these set-membership algorithms can also be interpreted as variable step size algorithms. Because of combining the advantages of the INSAF and SM-NSAF algorithms, both proposed SM-INSAF and SSM-INSAF algorithms can obtain better performance in low SNR scenarios. Importantly, the SSM-INSAF has smaller steady-state NMSD than the SM-INSAF. The reason behind this result is that the fluctuation of the step size computed by (26) (i.e., the SSM-INSAF) is smaller than that computed by (21) (i.e., the SM-INSAF), as shown in Fig. 5.

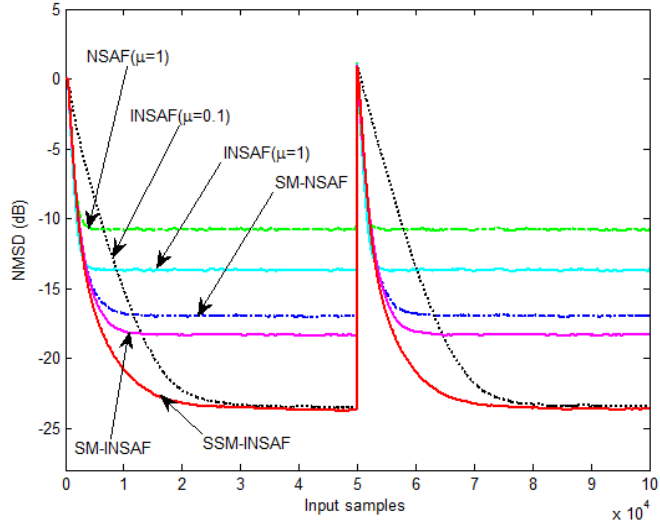

(a)

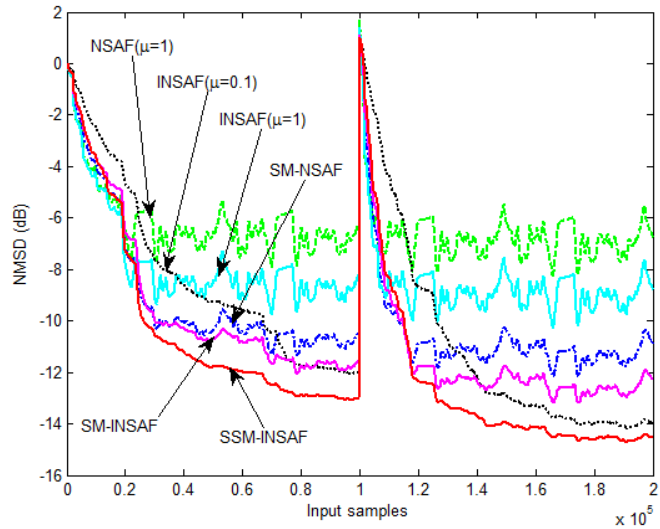

(b)

**Fig. 4.** The NMSD curves of various SAF algorithms for a dispersive echo path. (a) AR(1) input, (b) speech input. Parameters setting: $\gamma = \sqrt{3\sigma_\eta^2 / N}$ for the SM-NSAF; $t = 2$ for the SM-INSAF; $t = 0.75$, $\kappa = 1$ for the SSM-INSAF.

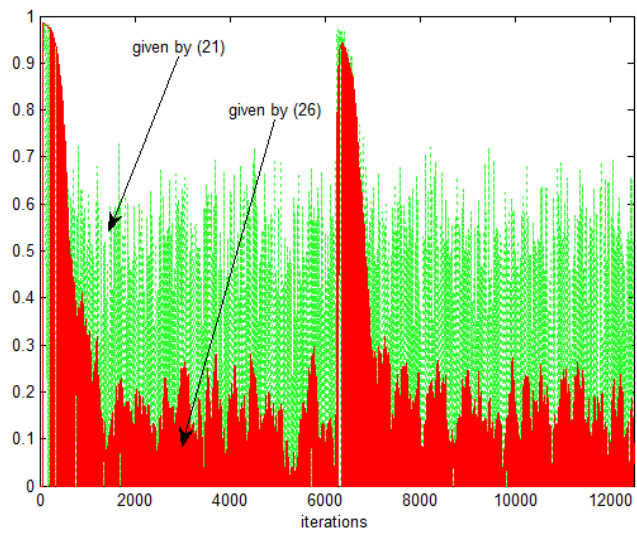

1515

**Fig. 5.** Variation of the step size $\mu_0(k)$ for the first subband. Simulation' setting is the same as Fig. 4 except for one trial.

Then, aiming to a sparse echo path given by Fig. 2(a), Fig. 6 provides a comparison of the SM-IPNSAF [20], IP-INSAF [21] and four proposed set-membership algorithms in the NMSD performance. For all the proportionate algorithms, the proportionate parameters are the same, i.e., $\lambda = 0$ and $\zeta = 0.0001$, and the regularization additive number is set to $\delta = \sigma_u^2/M$ when the input is speech signal. As one can see, the proposed SM-IP-INSAF and SSM-IP-INSAF algorithms have smaller steady-state NMSD than the SM-IPNSAF and IP-INSAF algorithms under the same convergence rate, and that of the SSM-IP-INSAF is the smallest among these algorithms. As compared to the SM-INSAF and SSM-INSAF algorithms, their proportionate versions (i.e., the SM-IP-INSAF and SSM-IP-INSAF algorithms) have faster convergence rate and better tracking capability for estimating a sparse echo path.

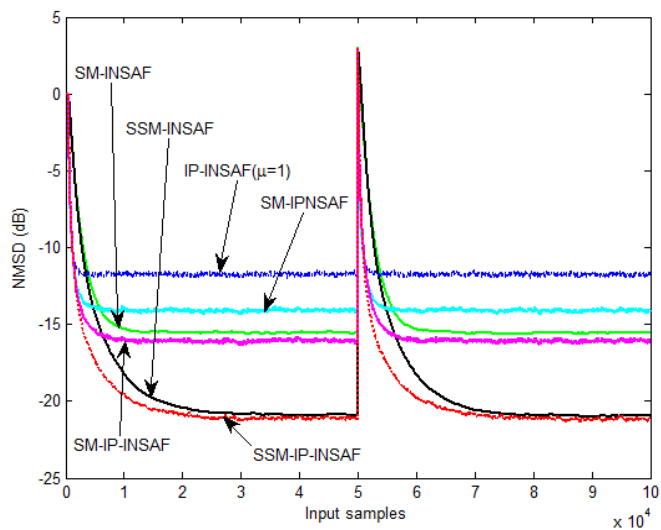

(a)

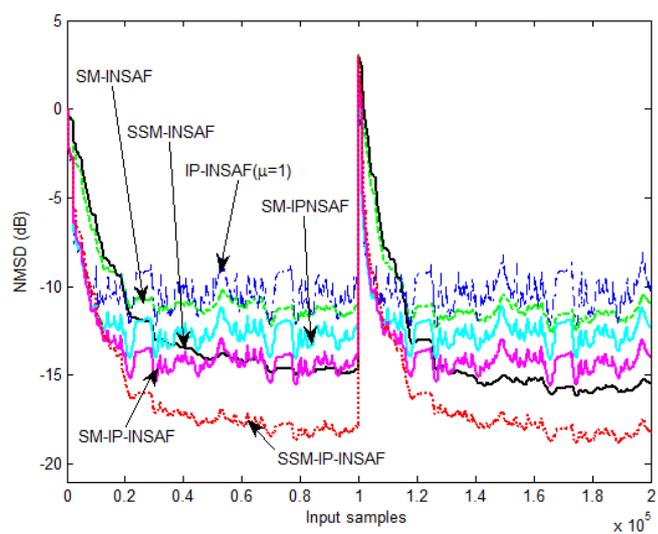

(b)

**Fig. 6.** The NMSD curves of various SAF algorithms for a sparse echo path. (a) AR(1) input, (b) speech input. Parameters setting: $\gamma = \sqrt{2\sigma_\eta^2 / N}$ for the SM-IPNSAF; $t = 2$ for the SM-IP-INSAF; $t = 0.75$, $\kappa = 1$ for the SSM-IP-INSAF.

Table 3 gives the update rates of the proposed algorithms for each subband, denoted by $F_{\text{up},i}$ for $i = 0,1,...,N-1$. The simulation conditions and setting of parameters are the same as Fig. 4(a) and 6(a). $F_{\text{up},i}$ is measured by $F_{\text{up},i} = \left(\text{N}_{update,i} / \text{N}_{total,i}\right)$, where $\text{N}_{update,i}$ and $\text{N}_{total,i}$ are the number of updates and the total number of iterations, respectively, in the case of $1 \times 10^5$ input samples; and then the average update rate $F_{\text{up}}$ over all subbands can be computed by $F_{\text{up}} = (1/N)\sum_{i=0}^{N-1} F_{\text{up},i}$. Note that, in terms of the maximum computational load per iteration (i.e., when the update is performed), given by $C_{\text{up}}$ in Tables 1 and 2, the proposed algorithms only have a slight increase relative to their original versions (i.e., the INSAF and the IP-INSAF). However, in Table 3 we can see that the update rates of the proposed algorithms are much less than 1, so according to (32) they reduce significantly the overall computational complexity.

**Table 3.** The update rates of the proposed algorithms for each subband

| Algorithms | $F_{\text{up},0}$ | $F_{\text{up},1}$ | $F_{\text{up},2}$ | $F_{\text{up},3}$ | $F_{\text{up},4}$ | $F_{\text{up},5}$ | $F_{\text{up},6}$ | $F_{\text{up},7}$ | $F_{\text{up}}$ |
|---|---|---|---|---|---|---|---|---|---|
| INSAF, IP-INSAF | 1 | 1 | 1 | 1 | 1 | 1 | 1 | 1 | 1 |
| SM-INSAF | 0.35 | 0.31 | 0.29 | 0.29 | 0.29 | 0.28 | 0.28 | 0.27 | 0.295 |
| SSM-INSAF | 0.52 | 0.50 | 0.48 | 0.48 | 0.48 | 0.47 | 0.48 | 0.48 | 0.486 |
| SM-IP-INSAF | 0.43 | 0.31 | 0.28 | 0.27 | 0.27 | 0.27 | 0.27 | 0.26 | 0.295 |
| SSM-IP-INSAF | 0.54 | 0.49 | 0.47 | 0.47 | 0.47 | 0.46 | 0.46 | 0.46 | 0.478 |

Also, the NMSD performance of four proposed algorithms for a dispersive case given by Fig. 2(a) is compared; and the results are plotted in Fig. 7. Parameters' choice of these algorithms is the same as Fig. 6 except for the proportionate parameter $\lambda = -0.5$. As can be seen from this figure, even though the echo path is dispersive, the SM-IP-INSAF and SSM-IP-INSAF algorithms behave almost as well as the SM-INSAF and SSM-INSAF algorithms, respectively.

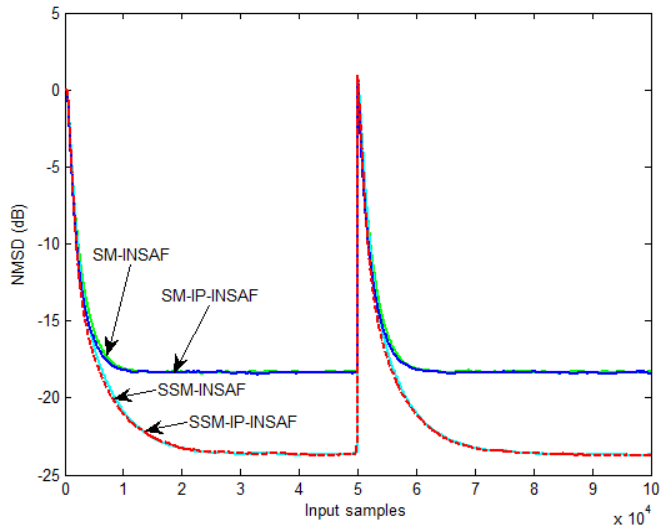

(a)

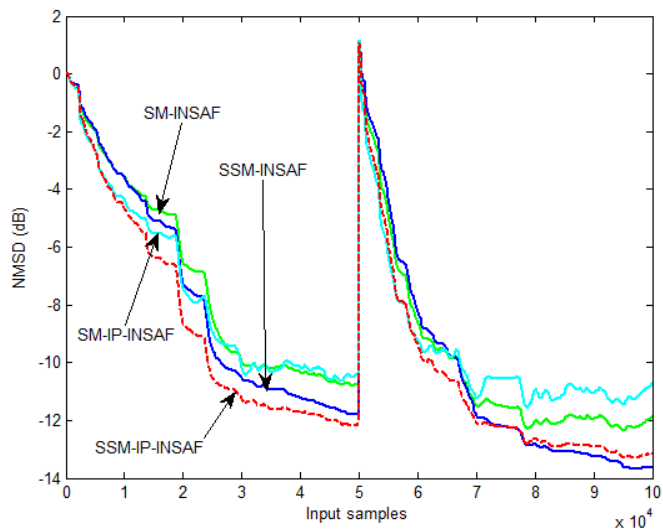

(b)

**Fig. 7.** Comparison of four proposed algorithms for a dispersive echo path. (a) AR(1) input, (b) speech input.

### 4.3. Comparisons with the fullband algorithms

In this example, we compare the performance of the proposed algorithms with that of two fullband AP (i.e., the SM-AP [23] and the SM-PAPA [25]) algorithms using a speech signal as the input. To fairly compare, Fig. 8(a) shows the performance of the SM-AP with different projection orders $K$, SM-INSAF and SSM-INSAF algorithms for a dispersive echo path, and Fig. 8(b) shows the performance of the SM-PAPA with different $K$ values, SM-IP-INSAF and SSM-IP-INSAF algorithms for a sparse echo path. In the SM-AP and SM-PAPA algorithms, the threshold is set to $\gamma = \sqrt{3\sigma_\eta^2}$; all the proportionate algorithms have the same proportionate formula (6) and the corresponding parameters are set to $\lambda = 0$ and $\zeta = 0.0001$; parameters setting of the proposed algorithms is the same as the above Figs. 4 and 6. As can be seen, for the colored input signal, the AP and proposed algorithms converge quickly relative to the NLMS-type (i.e., the AP when $K$=1), owing mainly to their inherent decorrelating properties in the time domain and the subband domain, respectively. However, these AP algorithms require larger computational cost than the proposed algorithms, as shown Tables 1 and 2.

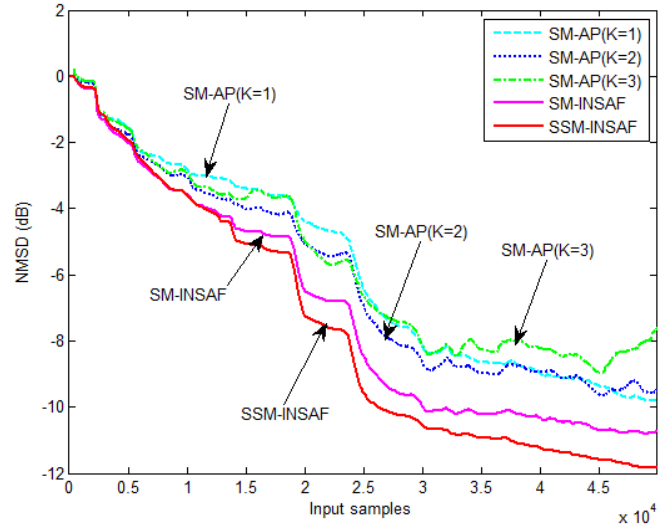

(a)

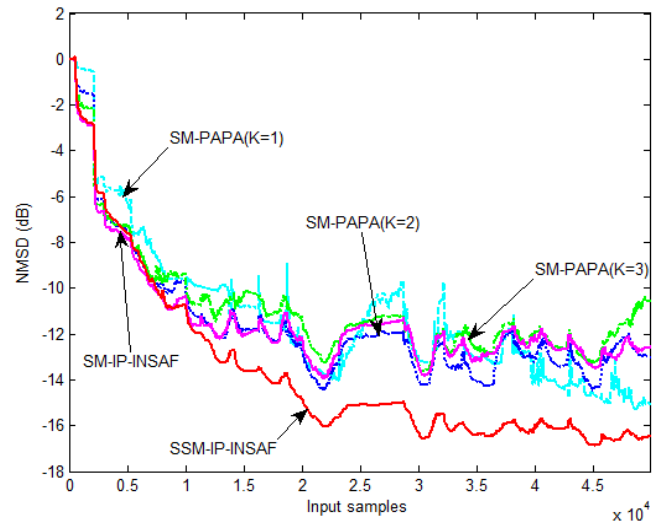

(b)

**Fig. 8.** The NMSD curves of the SM-IP-INSAF, SSM-IP-INSAF, SM-AP and SM-PAPA algorithms for a speech input. (a) dispersive echo path, (b) sparse echo path.

## 5. Conclusions

In this study, we derived the SM-INSAF and SM-IP-INSAF algorithms by incorporating the advantages of the set-membership filtering into the INSAF and IP-INSAF, respectively. Compared with the original INSAF and IP-INSAF algorithms, the proposed algorithms not only provide better steady-state performance under comparable convergence rate, but also save the overall computational cost. Next, their smooth versions, i.e., the SSM-INSAF and SSM-IP-INSAF, were proposed by smoothing subband error signals, further reducing the steady-state error. Moreover, the SM-IP-INSAF and SSM-IP-INSAF have faster convergence rate than the SM-INSAF and SSM-INSAF for sparse systems. Simulation results in an AEC scenario have demonstrated the superiority of our proposed algorithms for the case of low SNR.

**Acknowledgments**

This work was partially supported by National Science Foundation of P.R. China (Grant: 61571374, 61271340 and 61433011) and the work of Y. Yu was partially supported by China Scholarship Council Funding.

**Appendix**

In this appendix, we provide the convergence condition of the existing INSAF and IP-INSAF algorithms, which has not been studied in the previous literature. To obtain this purpose, we need the following assumptions:

*Assumption A1*: The background noise $\eta(n)$ is a white process with zero-mean and variance $\sigma_\eta^2$.

*Assumption A2*: The analysis filter banks $H_i(z)$, $i = 0,1,...,N-1$, are paraunitary, which was widely used in subband algorithms [30], [33], [34]. With this assumption, the decimated signal $d_{i,D}(k)$ can be expressed equivalently as:

$$d_{i,D}(k) = \mathbf{u}_i^T(k)\mathbf{w}_o + \eta_i(k) \tag{A1}$$

where $\eta_i(k)$, $i = 0,1,...,N-1$, denote the decimated subband noises which are generated by band-partitioning $\eta(n)$ through the analysis filter bank. Thus, according to assumptions A1 and A2, $\eta_i(k)$ is also zero-mean white process with variances $\sigma_{\eta_i}^2$.

*Assumption A3*: $\eta_i(k)$, $\mathbf{w}(l)$ for $l \leq k$, $\mathbf{u}_i(k)$ are statistically independent. It is well-known *independence assumption* in adaptive filtering algorithms [34], [35].

We start from the IP-INSAF since the INSAF is its a special case. Thus, we rewrite the update formula of the IP-INSAF as:

$$\mathbf{w}(k+1) = \sum_{p=0}^{P-1} b(p)\mathbf{w}(k-p) + \mu \sum_{i=0}^{N-1} \frac{\mathbf{G}(k)\mathbf{u}_i(k)\varepsilon_{i,\text{D}}(k)}{\mathbf{u}_i^T(k)\mathbf{G}(k)\mathbf{u}_i(k)}, \tag{A2}$$

$$\varepsilon_{i,\text{D}}(k) = d_{i,\text{D}}(k) - \mathbf{u}_i^T(k)\sum_{p=0}^{P-1} b(p)\mathbf{w}(k-p), \tag{A3}$$

where $b(p) = \alpha\rho^p$, $p = 0,1,...,P-1$, with $\sum_{p=0}^{P-1} b(p) = 1$. Subtracting (A2) from $\mathbf{w}_o$, we have

$$\tilde{\mathbf{w}}(k+1) = \sum_{p=0}^{P-1} b(p)\tilde{\mathbf{w}}(k-p) - \mu \sum_{i=0}^{N-1} \frac{\mathbf{G}(k)\mathbf{u}_i(k)\varepsilon_{i,\text{D}}(k)}{\mathbf{u}_i^T(k)\mathbf{G}(k)\mathbf{u}_i(k)}, \tag{A4}$$

where $\tilde{\mathbf{w}}(k) = \mathbf{w}_o - \mathbf{w}(k)$ denotes the weight error vector. Pre-multiplying both sides of (A4) by $\mathbf{u}_i^T(k)$, and using assumption A2 as well as the *diagonal assumption* in Section 3.3, we get:

$$\begin{aligned}\mathbf{u}_i^T(k)\tilde{\mathbf{w}}(k+1) &= \mathbf{u}_i^T(k)\sum_{p=0}^{P-1}b(p)\tilde{\mathbf{w}}(k-p) - \mu\varepsilon_{i,\text{D}}(k) \\ &= (1-\mu)\sum_{p=0}^{P-1}b(p)\mathbf{u}_i^T(k)\tilde{\mathbf{w}}(k-p) - \mu\eta_i(k)\end{aligned} \quad (A5)$$

Defining the *a posterior* subband error $e_{p,i}(k) = \mathbf{u}_i^T(k)\tilde{\mathbf{w}}(k+1)$ and the *a priori* subband error $e_{a,i}(k,p) = \mathbf{u}_i^T(k)\tilde{\mathbf{w}}(k-p)$, and then (A5) becomes:

$$e_{p,i}(k) = (1-\mu)\sum_{p=0}^{P-1}b(p)e_{a,i}(k,p) - \mu\eta_i(k). \quad (A6)$$

Under assumption A3, we can take the expectation of the squared of both sides of (A6) as

$$E\{e_{p,i}^2(k)\} = (1-\mu)^2\sum_{p=0}^{P-1}\sum_{q=0}^{P-1}b(p)b(q)E\{e_{a,i}(k,p)e_{a,i}(k,q)\} + \mu^2\sigma_{\eta_i}^2. \quad (A7)$$

Using again assumption A3, we obtain

$$E\{e_{a,i}(k,p)e_{a,i}(k,q)\} = E\{\tilde{\mathbf{w}}^T(k-p)\mathbf{R}_i\tilde{\mathbf{w}}(k-q)\}, \quad (A8)$$

where $\mathbf{R}_i = E\{\mathbf{u}_i(k)\mathbf{u}_i^T(k)\}$ is the covariance matrix of the $i^{\text{th}}$ subband input signal. According to Lemma 1 in [34], we know that $c\mathbf{I} \leq \mathbf{R}_i \leq d\mathbf{I}$, $0 < c \leq d < \infty$, where $\mathbf{I}$ denotes the identity matrix. It is assumed that the algorithm is convergent, we have the following relations:

$$E\{e_{p,i}(k)e_{p,i}(k)\} < E\{e_{a,i}(k,0)e_{a,i}(k,0)\}, \quad (A9)$$

$$E\{e_{a,i}^2(k,0)\} < \sum_{p=0}^{P-1}\sum_{q=0}^{P-1}b(p)b(q)E\{e_{a,i}(k,p)e_{a,i}(k,q)\} < E\{e_{a,i}^2(k,P-1)\}. \quad (A10)$$

So, to ensure (A9) and (A10) for all subbands, the term $(1-\mu)^2$ in (A7) must fulfill the condition $(1-\mu)^2 < 1$ which leads to

$$0 < \mu < 2. \quad (A11)$$

The relation (A11) is the convergence condition for the INSAF and IP-INSAF algorithms.